\documentclass[twocolumn,notitlepage,prx,superscriptaddress,longbibliography]{revtex4-2}
\usepackage{amsfonts}
\usepackage{changes}

\usepackage{textcomp}
\usepackage{times}
\usepackage{graphicx}
\usepackage{float}
\usepackage{latexsym,amsmath,amssymb,bm,euscript}
\usepackage{color}
\usepackage{xcolor}

\usepackage{changes}
\usepackage{subfigure}
\usepackage{epstopdf}
\usepackage[colorlinks=true,linkcolor=blue,citecolor=blue,urlcolor=blue]{hyperref}
\usepackage{soul}
\usepackage[normalem]{ulem}
\usepackage{mathrsfs}
\usepackage{amsmath}
\usepackage{amstext}
\usepackage[bottom]{footmisc}
\usepackage{amsbsy}
\usepackage{ulem}

\usepackage{MnSymbol}
\usepackage{physics}
\usepackage{graphicx}% Include figure files
\usepackage{dcolumn}% Align table columns on decimal point
\usepackage{bm}% bold math

\begin{document}

\title{{Global  Phase Diagram of D-wave Superconductivity in the Square-Lattice $t\text{-}J$ Model}}

\author{Feng Chen}
\affiliation{Department of Physics and Astronomy, California State University, Northridge, California 91330, USA}%

\author{F. D. M. Haldane}
\affiliation{Department of Physics, Princeton University, Princeton NJ 08544, USA}

\author{D. N. Sheng}
\affiliation{Department of Physics and Astronomy, California State University, Northridge, California 91330, USA}%

\date{\today}
\begin{abstract}
The Hubbard  and  closely related $t\text-J$ models are exciting platforms for  unconventional superconductivity (SC). Through state-of-the-art density matrix renormalization group calculations using the grand canonical ensemble, we address open issues regarding  the ground-state phase diagram of  the extended $t\text{-}J$ model on a square lattice in the parameter regime  relevant to cuprate superconductors. On large 8-leg cylinders, we demonstrate  that the pure $t\text{-}J$ model with only nearest-neighbor hoppings and superexchange interactions, for a wide range of doping 
($\delta=0.1-0.2$), hosts {robust d-wave superconductivity possibly coexisting with weak unidirectional pair density wave}. Furthermore, a small next nearest neighbor hopping $t_2$ suppresses {pair and charge density waves}%{the  inhomogeneous orders}{pair density wave and other co-coexisting orders} 
, resulting in a uniform d-wave SC phase in both electron- and hole-doped cuprate model systems. %\deleted{These results reveal a simple mechanism for SC in the $t\text{-}J$ model, in which the nearest neighbor hopping plays an essential role in driving the formation of Cooper pairs with real-space sign oscillations balancing the competition between the kinetic and exchange energies.}  
   Our work  validates the $t\text-J$ model as a proper minimum model for {the emergence of superconductivity in } cuprate superconductors.
\end{abstract}
\maketitle
\section{Introduction} The mechanism of high-temperature superconductivity (SC) in cuprate compounds is one of the major challenges in condensed matter physics~\cite{lee_doping_2006,Keimer_Nature_2015,Proust_ARCMP_2019}.
The emergence of superconductivity from doping the antiferromagnetic Mott insulating phase of cuprates
suggests the square-lattice  Hubbard model in the strong coupling regime and the closely related  $t\text{-}J$ model as the minimal models for its theoretical understandings~\cite{anderson_resonating_1987, zhang_effective_1988,Rokhsar_PRL_1988,Keimer_Nature_2015,Proust_ARCMP_2019,Anderson_2004,lee_doping_2006,Sachdev_RMP_2003,weng_phase_1997,Masao_RPP_2008}.  {The investigation of the existence of SC and the pairing mechanism in these models with purely repulsive interactions, as opposed to phonon-mediated electron attractions in the conventional superconductors, has challenged physicists for decades and led to the development of many theoretical concepts and numerical methods~\cite{arovas_hubbard_2022,qin_hubbard_2022}.}
 On the analytical side, the resonating valence bond (RVB) theory has presented a picture for the development of unconventional SC from a spin-liquid background~\cite{anderson_resonating_1987,anderson_resonating--valence-bond_1987,kivelson_topology_1987,Liang_1988, Anderson_2004, lee_doping_2006, scalapino_common_2012},  {while the spin-fluctuation mediated attraction~\cite{scalapino_common_2012} offers an alternative perspective}. New concepts such as spin-charge separation, spin liquid, electron fractionalization, and intertwined spin and charge orders have been extensively explored and may play important roles in understanding doped Mott insulators~\cite{anderson_resonating_1987,Rokhsar_PRL_1988,Sachdev_RMP_2003,weng_phase_1997, wen1991,Senthil_2000,lee_doping_2006,Kaul_NP_2007,Balents2007dual,fradkin_colloquium_2015,weng_superconducting_2011,Jiang_stripe_2022}.  However, various analytical approximations are not controlled in identifying the quantum phases of these strongly correlated systems.
On the numerical side, large-scale simulations in recent years have often identified  the stripe phase in the ground states of the pure $t\text{-}J$ and  Hubbard models with only nearest-neighbor (NN) hoppings~\cite{white_density_1998,White_PRL_2003, Hager_PRB_2005,Berg_2009,dodaro_intertwined_2017,Garnet_Science_2017,Ido_PRB_2018,Jiang_Science_2019, ponsioen_period_2019,simons_absence_2020,jiang_ground-state_2021,lu_sign_2023,jiang_ground-state_2024}.
Meanwhile, there are also indications that the stripe phase has local pairing tendency~\cite{jiang_ground-state_2021, lu_sign_2023,wietek_fragmented_2022}
toward developing SC %%and is in strong competition with SC 
order~\cite{corboz_competing_2014,Raczkowski_unidirectional_2007,yang_nature_2009,ponsioen_period_2019,li_study_2021,lu_emergent_2024}. Therefore, it is fundamentally important to re-examine if the pure t-J model can host SC in its ground state, and identify the physical reasons for its emergence or absence.

For a more realistic modeling of cuprate systems, the extended $t\text-J$ %\deleted{($t_1\text-t_2\text-J_1(\text-J_2)$)} 
 and Hubbard models with both NN and next-nearest-neighbor (NNN)  hoppings ($t_1, t_2$)  have been extensively studied and a positive or negative $t_2$ is adopted for electron- or hole-doped  cuprates to reproduce the Fermi surface in experiments~\cite{jiang_ground-state_2021,Damascelli_RMP_2003,Nazarenko_photoemission_1995,Kim_systematics_1998}.
Earlier numerical studies have found that positive $t_2$ enhances d-wave SC order whereas negative $t_2$ promotes the stripe phase and suppresses SC~\cite{white_competition_1999,martins_qualitative_2001,himeda_stripe_2002}, or favors plaquette-type pairing in width-4 cylinders~\cite{dodaro_intertwined_2017,Jiang_Science_2019,simons_plaquette_2020,jiang_ground_2020}. % around the optimal hole doping.
More  recently,  density matrix renormalization group (DMRG)~\cite{white_density_1992} studies of the extended $t\text-J$ model on wide cylinders with the width $L_y$ upto $L_y=8$ have reached a consensus on the presence of robust d-wave SC  for $t_2/t_1\gtrsim 0.1$, marking a significant step towards establishing d-wave SC for electron-doped cuprate models~\cite{gong_robust_2021,jiang_ground-state_2021,lu_sign_2023,lu_emergent_2024,shen2024groundstateelectrondopedttj} or doped spin liquid (or spin-liquid-like states)~\cite{jiang_high_2021,jiang_superconducting_2023}. However, 
numerical results on the hole-doped  side ($t_2<0$) remain controversial. Previous DMRG studies 
reported the spin and charge stripes and absence of SC in the ground state of the hole-doped $t\text{-}J$ model~\cite{jiang_ground-state_2021,jiang_pairing_2022,lu_sign_2023}, 
whereas latest DMRG simulations using  larger bond dimensions have identified the melting of charge density wave (CDW) and emergence of weak
SC as the system widens from $L_y=6$ to $8$  at $1/8$ hole doping~\cite{lu_emergent_2024}. 
 It remains  challenging to  fully understand  the  competition and interplay between
  CDW, magnetic order and SC {, and unify the emergence of SC in the family of $t\text-J$ models}.

\begin{figure}[htbp]
\includegraphics[width=1\linewidth]{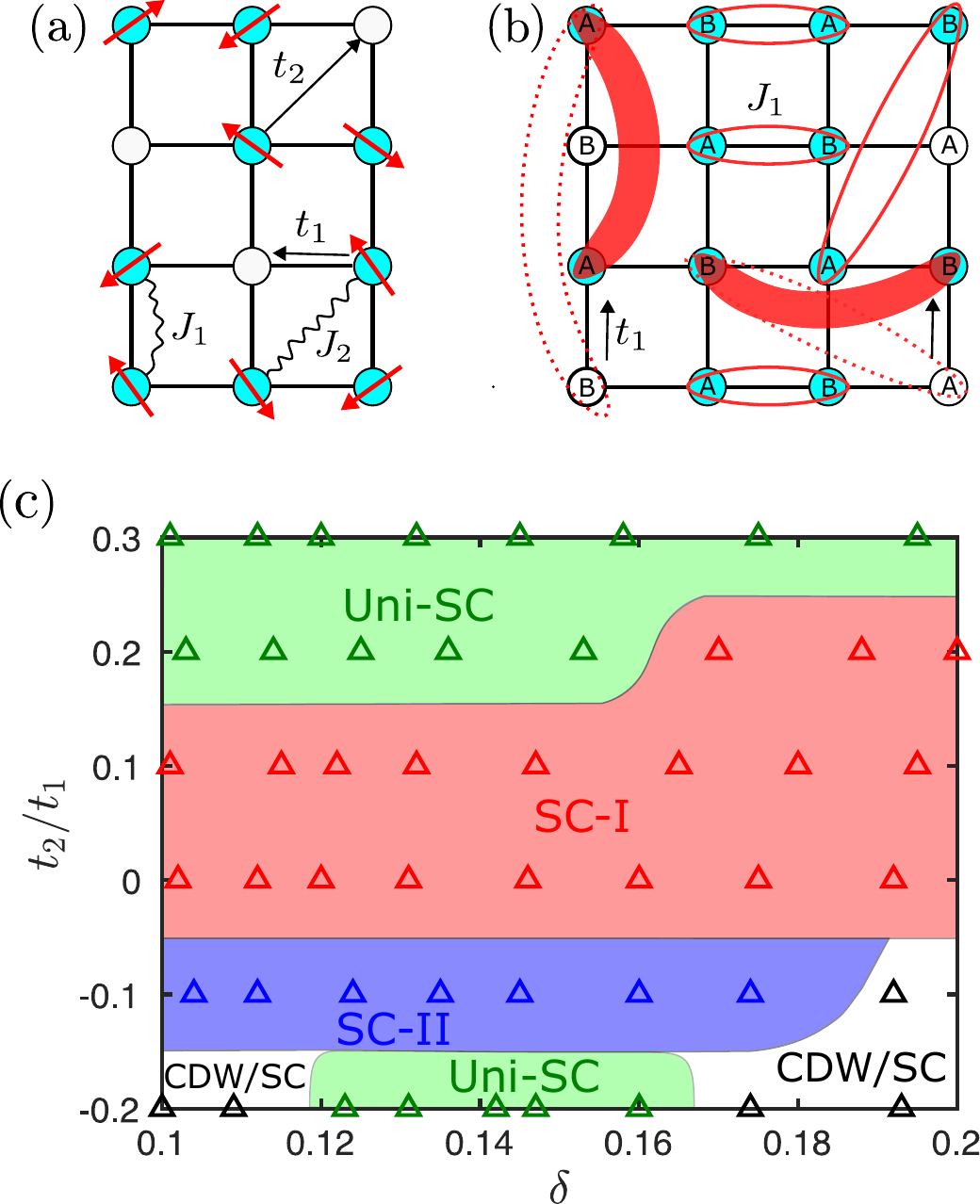}
\caption{(a) Illustration of the extended $t{\text -}J$ model. (b) Illustration of the development of same-sublattice pairings (denoted by filled ovals) from opposite-sublattice ones (denoted by unfilled ovals) in the pure $t{\text -}J$ model as a consequence of NN electron hopping. (c) Ground-state phase diagram under the variation of doping level $\delta$ and ratio $t_2/t_1$. Three SC phases are found.  {The uniform d-wave superconducting phase is denoted by Uni-SC, the d-wave SC phase coexisting with short-range spin and charge stripes and potentially weak period-2 PDW is called SC\text-I, and the d-wave SC phase with coexisting weak long-range CDW is denoted by SC\text-II.}  The white regimes at the lower corners labeled by CDW/SC have strong competition between  CDW and SC, which cannot be fully resolved with the currently accessible bond dimensions.  {All phases are characterized by short-range spin correlations.}}
\label{Fig_phase_diagram}
\end{figure}

In this article, we address open issues regarding the emergence of unconventional SC in the extended $t\text-J$ model on a square lattice % $t_1{\text -}t_2{\text -}J_1{\text -}J_2$ model  
 through state-of-the-art  DMRG  simulations of  grand canonical~\cite{jiang_ground-state_2021} systems with $L_y=8$. We establish a global quantum phase diagram (Fig.~\ref{Fig_phase_diagram}(c)) where $d$-wave SC order is prevalent over the entire parameter regime with the doping level $0.1\leq \delta \leq 0.2$ and  hopping ratio  $-0.2\leq t_2/t_1\leq 0.3$. 
Particularly, robust d-wave SC is identified in the ground state of the pure $t\text{-}J$ model ($t_2=0$), {possibly}
coexisting with weak period-two unidirectional pair density wave (PDW)~\cite{agterberg_physics_2020}. In the hole-doped $t\text-J$ model with negative $t_2/t_1\sim-0.1$,  d-wave SC coexists with weak long-range CDW.  
A common uniform SC phase emerges at relatively larger magnitudes of $t_2/t_1\sim\pm0.2$ by suppressing  PDW and CDW, suggesting a symmetry between the electron- and hole-doped $t\text-J$ models.
 {The universal emergence of SC in the extended $t\text{-}J$ model,  {especially at $t_2=0$}, suggests the important role %to be driven by 
of the frustrating NN hopping ($t_1$) that turns the  spin singlets into coherent Cooper pairs upon doping.}
Our results validate the extended  $t\text-J$ model as a suitable minimal model for the emergence of SC in cuprate and provide benchmark results for future theoretical studies of its mechanism.

\section{Model}
We consider the square-lattice $t_1\text{-}t_2\text{-}J_1\text{-}J_2$ model defined as 
\begin{equation}
\label{H-tJ}
\begin{split}
  H = &-\sum_{\{ij\},\sigma}t_{ij}(\hat{c}^{\dagger}_{i,\sigma} \hat{c}_{j,\sigma} + {\rm H.c.})  \\
  &+ \sum_{\{ij\}} J_{ij} (\hat{\bf S}_i \cdot \hat{\bf S}_j - \frac{1}{4} \hat{n}_i \hat{n}_j)-\mu\sum_i\hat{n}_i,
  \end{split}
\end{equation}
where ${\hat{c} }_{i\sigma }^{\dagger }$ and ${\hat{c} }_{i\sigma }$ are the creation and annihilation operators of the electron with spin $\sigma$ ($\sigma = \pm 1/2$) at site $i=(x,y)$, ${\hat{\bf S} }_i$ is the spin-$1/2$ operator and ${\hat{n} }_i =\sum_{\sigma } {\hat{c} }_{i\sigma }^{\dagger } {\hat{c} }_{i\sigma }$ is the electron number operator. Double occupancy is prohibited. The electron hoppings $t_{ij}$ and the spin exchange interactions $J_{ij}$ are restricted to the NN ($t_1$, $J_1$) and NNN ($t_2$, $J_2$) terms. % (as $t_{1,2}$ and $J_{1,2}$). 
We consider  a cylinder that is periodic along the circumferential ($\hat{y}$) direction and open along the axial ($\hat{x}$) direction with width  $L_y$ and length $L_x$ respectively. The total number of  sites is $N = L_x \times L_y$.  Finally, we set $J_1=1$ as the energy unit and $t_1= 3$, corresponding to a Hubbard model at $U/t_1=12$~\cite{jiang_high_2021, gong_robust_2021}, and vary $t_2$ and $J_2$ together via the relation $J_2/J_1=(t_2/t_1)^2$.

\section{Phase Diagram}
We obtain a phase diagram for $L_y=8$ systems in the regime of doping level $0.1\leq \delta \leq 0.2$ and hopping ratio $-0.2\leq t_2/t_1\leq 0.3$ as  shown in Fig.~\ref{Fig_phase_diagram}(c). 
We identify  three different SC phases: (1) Uniform $d$-wave SC at both $t_2/t_1\gtrsim 0.15$ and $t_2/t_1\lesssim -0.15$, referred to as Uni-SC; (2) D-wave SC coexisting with   {short-range spin and charge stripes} and potentially weak period-2 unidirectional PDW order
%with pair momentum $\boldsymbol{k}_\text{PDW}=(0,\pi)$   
for  $-0.05\lesssim t_2/t_1 \lesssim 0.15-0.25$ depending on the doping level, referred to as SC-I; (3) D-wave SC coexisting with weak long-range CDW order at $-0.15\lesssim t_2/t_1\lesssim-0.05$, referred to as SC-II. In the bottom corners  of the phase diagram with $t_2/t_1\sim -0.2$ and $\delta\sim 0.11$
and $0.19$, there is also a possible CDW  phase coexisting with weaker or vanishing SC order. 
A common feature of all SC phases is the development of same-sublattice spin-singlet Cooper pairs induced by NN hopping  as illustrated in Fig.~\ref{Fig_phase_diagram}(b).

\begin{figure*}[!htbp]
\includegraphics[width=1\linewidth]{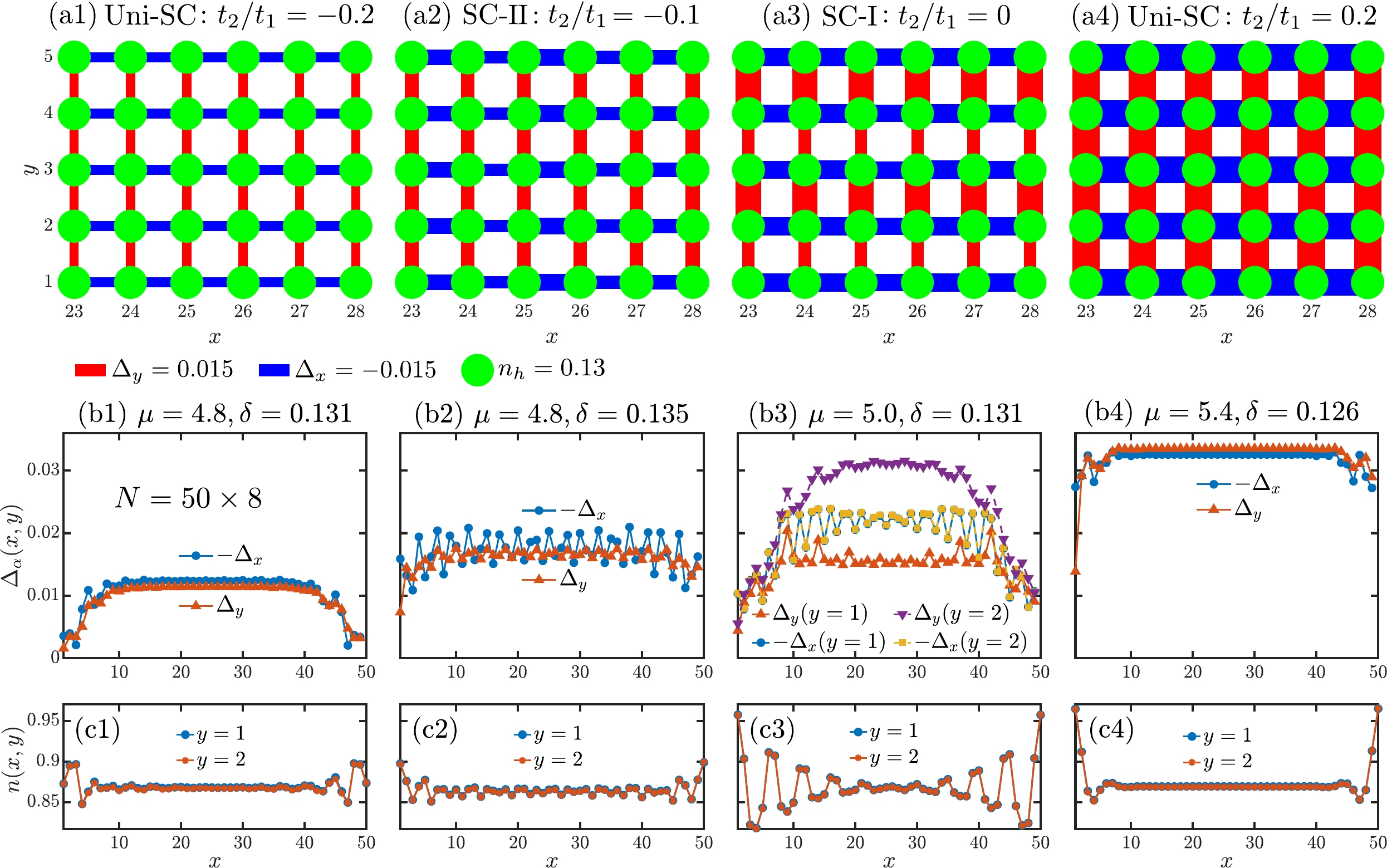}
\caption{The SC pairing orders and electron density profiles for four representative $t_2/t_1$ around 1/8 doping. (a) 2D plots for the NN pairing orders $\Delta_{\alpha=x,y}(\boldsymbol{r})$ and hole density $n_h(\boldsymbol{r})=1-\langle \hat{n}_{\bf r} \rangle$ obtained at $M=12000$. The bond thickness is proportional to the pairing amplitude and blue/red color denotes positive/negative sign respectively. The area of the green circle is proportional to the hole density $n_h(\boldsymbol{r})$ at each site. 
(b) The middle panels show the NN pairing orders $\Delta_{\alpha=x,y}$ along $x$-direction, which are uniform along $\hat{y}$ for (b1, b2) and (b4) but oscillating along $\hat{y}$ with a period of two lattice constants for $\Delta_{y}$  in (b3), i.e. in the pure $t\text{-}J$ model. (c) The lower panels show the electron density profiles $n(x,y)$ for $y=1,2$.}
\label{SC_CDW}
\end{figure*}

 To identify the SC order, we  examine the dominant spin-singlet NN pairing order defined as:
\begin{equation}
\Delta_\alpha(x,y)
=\langle\hat{c}_{(x,y),\uparrow}\hat{c}_{(x,y)+\hat{\alpha},\downarrow}-\hat{c}_{(x,y),\downarrow}\hat{c}_{(x,y)+\hat{\alpha},\uparrow} \rangle/\sqrt{2},
\label{Pairing_order}
\end{equation}
where $\hat{\alpha}=\hat{x}$ or $\hat{y}$ denotes the direction of the NN bond.
%\deleted{We first demonstrate the global emergence of SC as illustrated in Fig.~\ref{SC_CDW}.}
For four representative parameter points with $\delta$ around 1/8, $t_2/t_1$ from -0.2 to 0.2 and a large system size $N=50\times 8$, we show NN SC pairing orders $\Delta_{x,y}(\boldsymbol{r})$ %with a magnitude proportional to the bond width 
in Fig.~\ref{SC_CDW} (a1-a4). The opposite signs for x- and y-bonds demonstrate a robust d-wave symmetry.
Furthermore,  $\Delta_x$ and $\Delta_y$ are nearly constant in the bulk for the Uni-SC phase (Fig.~\ref{SC_CDW} (b1,b4)), while
there are noticeable oscillations of $\Delta_x$ along $x-$direction in both
the SC-I and SC-II phases (Fig.~\ref{SC_CDW} (b2,b3)), indicating the coupling between CDW (Fig.~\ref{SC_CDW} (c2,c3)) and the SC order $\Delta_x$~\cite{agterberg_physics_2020,fradkin_colloquium_2015}.
In the SC-I phase for the pure $t\text{-}J$ model, $\Delta_y$ is oscillating with a period of 2 lattice spacings
along $y-$direction as shown in Fig.~\ref{SC_CDW} (a3), corresponding to a PDW component with a pair momentum $\boldsymbol{k}_\text{PDW}=(0,\pi)$.

The electron density $n(x,y)$ generally preserves translational invariance along the $y$-direction %and $n(x,y)$ for $y=1, 2$ are shown 
as demonstrated in Fig.~\ref{SC_CDW}(c1-c4).  We identify  a very uniform  $n(x,y)$ in 
the Uni-SC phase at $t_2/t_1=\pm 0.2$ (see Fig.~\ref{SC_CDW} (c1, c4)), and a weak long-range CDW order along $x$ direction in the SC-II phase at $t_2/t_1=-0.1$. For the pure $t\text{-}J$ model, stripe-like density patterns along the $x$-direction appear near the open boundaries (see Fig.~\ref{SC_CDW} (c3)) but become flattened toward the center of the system.

Overall, the SC order becomes stronger as $t_2$ becomes more positive, and both positive and negative $t_2$ of large enough magnitudes ($\abs{t_2/t_1}\gtrsim 0.15$) suppress CDW and PDW, resulting in a uniform d-wave SC phase.
The uniform SC phase at $t_2/t_1=0.2-0.3$ identified here is consistent with earlier results from  DMRG studies in either the GCE 
~\cite{jiang_ground-state_2021} or the CE~\cite{gong_robust_2021,lu_emergent_2024}, further establishing the robust SC phase in the electron-doped $t\text{-}J$ model.  More significantly, hole and electron doping are unified by the common uniform SC phase on both sides.

\section{Uniform SC Phase and Competing Stripe Order in the Hole Doped $t\text{-}J$ Model}
\label{hope-doped}
\begin{figure}[!htbp]
\includegraphics[width=1\linewidth]{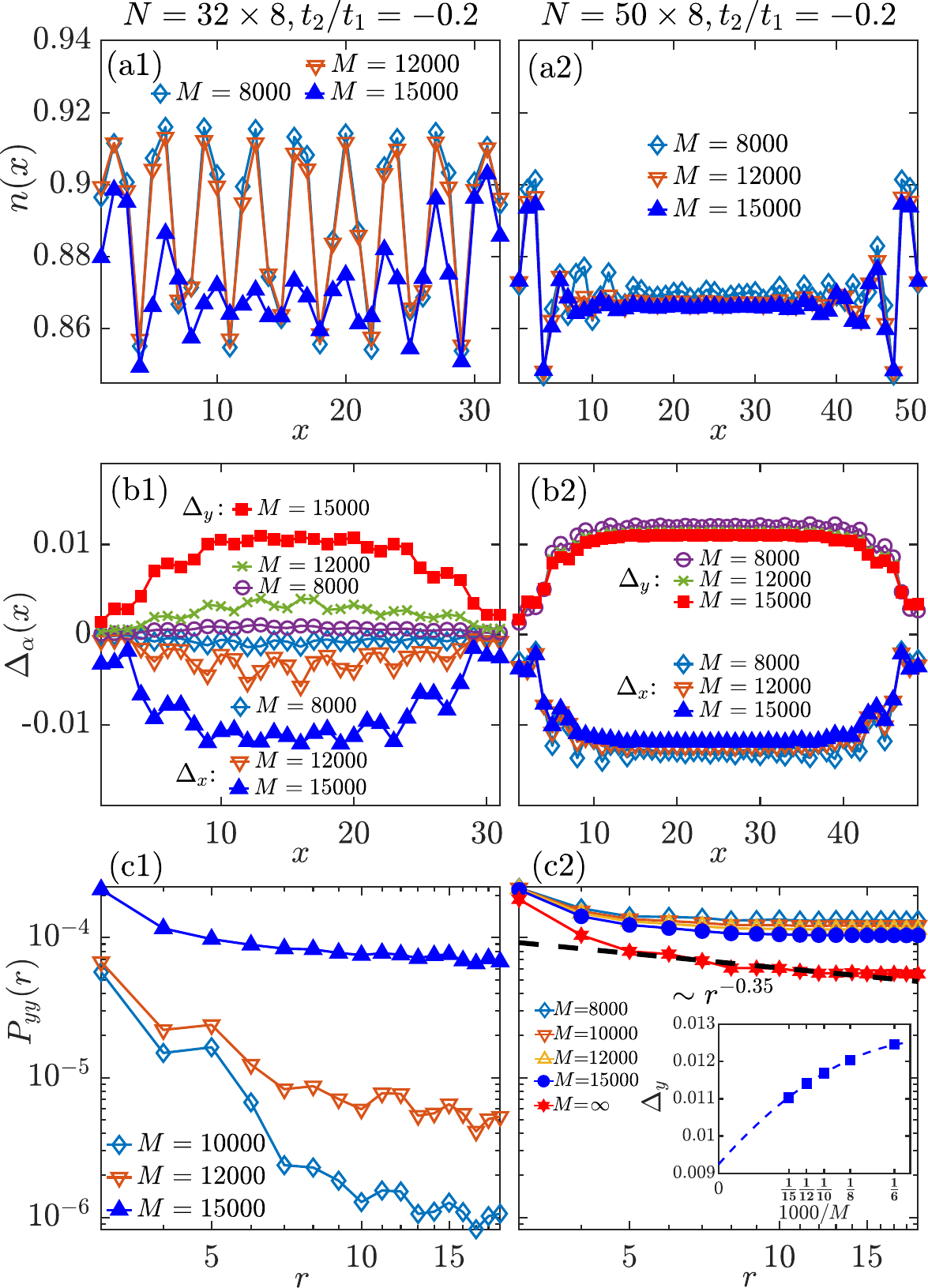}
\caption{ {The evolution of SC and CDW with the cylinder length and bond dimension} for the hole-doped $t\text-J$ model with $t_2/t_1=-0.2$ and $\mu=4.8$. 
(a1-2) Electron density profiles, (b1-2) pairing orders and (c1-2) pair-pair correlations at different bond dimensions for system sizes $N=32\times 8$ (left panels) and $N=50\times 8$ (right panels).  The inset in (c2) shows the pairing order for $N=50\times 8$ at infinite $M$ through second-order polynomial extrapolation, giving $\Delta_y(M=\infty)\approx 0.009$. { The extrapolated pair-pair correlation shows a power-law decay exponent around 0.35.}}
\label{tm02}
\end{figure}

A CDW in the stripe form  was  identified either as the  ground state~\cite{jiang_ground-state_2021} or a weak order  co-existing with power-law  SC correlations~\cite{lu_emergent_2024}  for the hole-doped $t\text{-}J$ model. 
Here we demonstrate how SC orders emerge through suppressing the stripe order in large 8-leg systems. 
Fig.~\ref{tm02} shows the charge densities, pairing orders $\Delta_{x,y}$ and pairing correlations $P_{yy}(r)=\langle \hat{\Delta}_y(x_0,y_0)\hat{\Delta}_y(x_0+r,y_0)\rangle$ in 8-leg cylinders of different lengths at $t_2/t_1=-0.2$. % and $\mu=4.8$
%corresponding to an average doping $\delta\sim 0.13$. 

In Fig.~\ref{tm02}(a1) for $N=32\times8$, charge stripes are quite robust until
$M$ reaches the large value of $M=15000$, where they are substantially weakened.
The SC orders $\Delta_{x,y}$ gradually develop from being nearly vanished to  notably strong when $M$ is increased from 8000 to 15000 as shown in Fig.~\ref{tm02}(b1). Consistently, the pairing correlation $P_{yy}(r)$ in Fig.~\ref{tm02}(c1) also increases with $M$ and exhibits a robust quasi-long-range SC order at $M=15000$~\cite{mermin_absence_1966}. {The melting of CDW and the emergence of SC arise from two factors: firstly, the CDW order is more classical and is the optimal low-energy state when the bond dimension is relatively small. Upon increasing the bond dimension, the  more entangled state such as a coherent SC state can be developed. This is clearly demonstrated by the rise of SC order, the weakening of CDW, and the near invariance of average electron density ($\delta\sim 0.11$) when $M$ is increased from 6000 to 12000 as shown in Fig.~S3~\cite{SM}; secondly, when $M$ is increased from $12000$ to $15000$, more holes are attracted into the system due to the gaining of negative  hole hopping energy dominating the loss of exchange energy. 
%removal of high-energy electrons at the top of the band~\cite{Energy_change}. 
This increase of hole doping promotes SC over CDW as shown by the CDW/SC to Uni-SC transition in the lower-left corner of Fig.~\ref{Fig_phase_diagram}(c). Eventually, the SC phase is clearly reached at $\delta\sim 0.13$ and $M=15000$.}

In a longer system with $N=50\times8$,  the SC orders $\Delta_{x,y}$ in Fig.~\ref{tm02}(b2) are already quite robust at $M=8000$ and are nearly independent 
of the bond dimension for $M=8000-15000$, indicating a more robust SC phase for larger systems (see also Fig.~S1(a)~\cite{SM}). {Meanwhile, the electron density is nearly uniform in the bulk, in contrast with the much stronger density oscillations at a similar doping for $N=32\times 8$ (see Fig.~\ref{tm02}(a1) with $M=15000$)}.     
The SC order is weaker at the boundary but grows rapidly into the bulk (Fig.~\ref{tm02}(b2)), whereas the charge modulations only appear around the boundaries (Fig.~\ref{tm02}(a2)), {further reflecting the competition between CDW and SC}.
%The uniform SC order and electron density in the bulk mark a uniform SC phase. 
{In Fig.~\ref{tm02}(c2), the SC order is extrapolated to infinite $M$ and gives $\Delta_y(M=\infty)\approx 0.009$, supporting the robustness of the SC order. Correspondingly, the extrapolated pair-pair correlation exhibits a power-law decay with a small exponent around 0.35, indicating a quasi-long-range SC order. On the other hand, the simulations in the CE according to Ref.~\cite{lu_emergent_2024} using bond dimensions $M_c$ up to 28000 yield an exponent around 1.64. The GCE simulations access symmetry-broken states and incorporate quantum fluctuations through $M$, whereas the CE simulations obtain symmetric states and allow long-range correlations to develop as $M_c$ increases. For sufficiently large $M$ and $M_c$, one would expect the same results from both ensembles as shown for six-leg systems in Fig.~S1~\cite{SM}. However, due to the limits of accessible ranges of $M$ and $M_c$, it is currently not feasible to determine the exact power-law decay exponent of pair-pair correlations in the eight-leg hole-doped system. Instead, we interpret the results from both ensembles as the upper and lower bounds for the exact exponent, i.e. it lies between 0.35 and 1.64, and take the finite value of $\Delta_\alpha$ (extrapolated to $M=\infty$) as clear evidence for SC.} 

\section{{Dominant D-wave SC and Other Weaker Orders } in the Pure $t{\text -}J$ Model}
Now we focus on the pure $t\text{-}J$ model with $t_2=J_2=0$.  
The pairing order $\Delta_y$ exhibits a double unit cell along the $y$-direction, and its average  $\bar{\Delta}_y(x) =[\Delta_y (x, y = 2) + \Delta_y (x, y = 1)] /2$    
is close in magnitude to $\Delta_x$ (Fig.~\ref{SC_CDW} (b3)), forming the d-wave SC background. The PDW order is obtained by subtracting the d-wave background  from the local SC order:
$\Delta_y^{\text{pdw}}(x,y)=[\Delta_y (x, y)-\bar{\Delta}_y (x)]$,  which has the following y-dependence:
\begin{equation}
     \Delta_y^{\text{pdw}}(x,y)  =(-1)^y\Delta_y^\text{P}(x), %\delta_{\hat{\boldsymbol{r}},\boldsymbol{e}_y},
    \label{Delta_P}
\end{equation}
indicating a unidirectional PDW with Cooper-pair momentum $\boldsymbol{k}_\text{PDW}=(0,\pi)$.
The magnitude of the PDW order, however, drops as $M$ increases and is extrapolated to  $\Delta_y^\text{P}(M=\infty) \approx 0.001$ as shown in Fig.~\ref{PDW}(b). {
%It therefore demonstrates the importance of finite bond dimension scaling. 
Although this is much weaker than the average d-wave SC order $\bar{\Delta}_y(M=\infty)\approx 0.019$ as shown in Fig.~\ref{PDW}(a), a significant proportion with $\Delta^P_y=27\% \bar{\Delta}_y$ is observed at $M=20000$ and a nonzero PDW component appears in a finite regime of the phase diagram. These suggest the possibility of its existence in the ground state.} 
We illustrate the pattern of the PDW order  $\Delta_y^{\text{pdw}}(x,y)$ in real space in Fig.~\ref{PDW}(c). To the best of our knowledge, the PDW order has only been  identified in more complex models for cuprate, e.g. the three-band models~\cite{jiang_pair_2023_arxiv,jiang_pair_2023_prb,ponsioen_superconducting_2023},
% it is highly desired 
or  explored in phenomenological  and mean-field studies~\cite{Berg_2009,choubey_atomic-scale_2020,tu_evolution_2019,agterberg_physics_2020}.  {The strong-weak period-2 pattern is also observed in both the spin and charge bond orders along y, indicating the intertwining between SC, spin and charge orders (see Fig. S5 in the SM~\cite{SM}).}

{The quasi-long-range d-wave SC order is further supported by CE simulations, which show power-law decays for $P_{yy}(r)$ and $P_{xx}(r)$ with exponents around 1.84 in Fig. \ref{PDW}(d). This power-law behavior can only be established by using extremely large bond dimensions up to $M_c=40000$. Earlier calculations~\cite{xu_coexistence_2024} at smaller bond dimensions $M_c\leq 28000$ found dominant CDW order and no clear evidence for SC.  In contrast, GCE simulations can detect SC reliably at smaller $M\sim 10000$, again demonstrating the numerical  {superiority} of GCE in identifying SC phases. 

 {We find that $P_{yy}(r)$ from the CE simulations are identical along different legs, in contrast with the oscillations of $\Delta_y$ along the $y$-direction observed in the GCE. Considering also the weakness of the extrapolated $\Delta^P_y(M=\infty)$, one should rely on larger bond dimension simulations in both ensembles to determine the existence of PDW in 8-leg systems. Moreover, the fact that the period-2 PDW appears only along the narrower $y$-direction %instead of the more extended $x$-direction 
raises the possibility that it might be a finite-size effect as the cylinder geometry breaks $C_4$ symmetry. Studies on wider systems utilizing the tensor network method are highly desired to pin down the fate of PDW in the 2D limit~\cite{zheng2024competingpairdensitywave}.}

\begin{figure}[!htbp]
\includegraphics[width=1\linewidth]{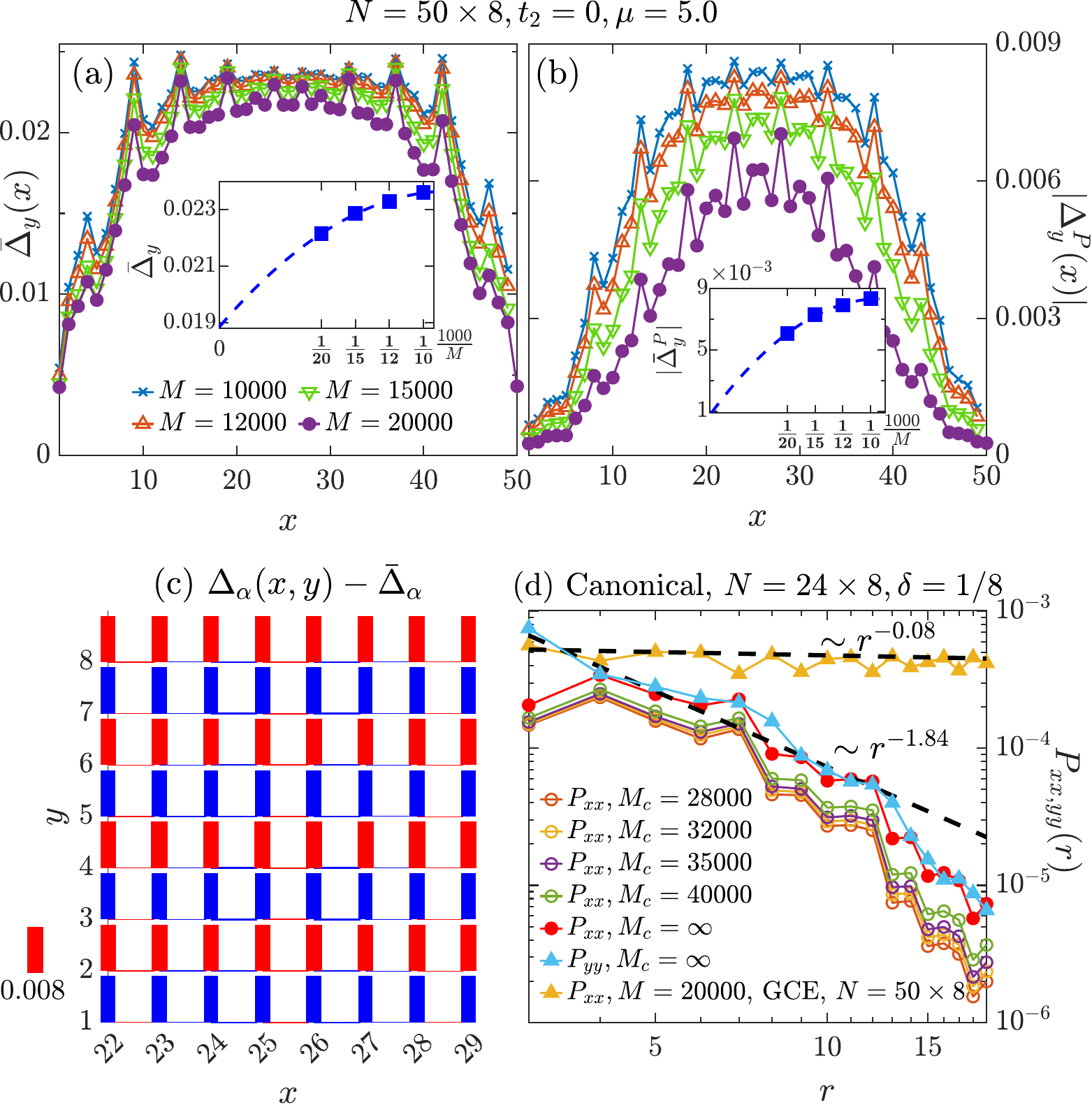}
\caption{D-wave SC {possibly coexists with weak} PDW in the SC-I phase at $t_2=0, \mu=5.0$ and $N=50\times8$. 
(a) The average $y$-bond pairing order $\bar{\Delta}_y(x)\equiv\left[\Delta_y(x,y=2)+\Delta_y(x,y=1)\right]/2$ with 
%The inset shows the second-polynomial extrapolation of the bulk $\bar{\Delta}_y$ versus $1/M$: 
$\bar{\Delta}_y(M=\infty)=0.019$. (b) The PDW order $\Delta^\text{P}_y(x)$ defined in Eq.~\ref{Delta_P}
%. The bulk value after extrapolation is 
with $\Delta_y^\text{P}(M=\infty)=0.001$. (c) Illustration of the unidirectional PDW component.
%with momentum $\boldsymbol{k}_\text{PDW}=(0,\pi)$. 
The $x$- and $y$-bond mean values $\bar{\Delta}_x=-0.022$ and $\bar{\Delta}_y=0.023$ are subtracted respectively.
%from $\Delta_x$ and $\Delta_y$ respectively. 
The red/blue color represents positive/negative values respectively and the bond widths are proportional to their magnitudes.
%of the quantity represented. A reference bond and its magnitude is put beside the plot. 
$M=12000$ is used. (d) $X$- and $Y$-bond pair-pair correlations for the pure t-J model in the CE at $\delta=1/8$ and $N=24\times 8$, which decay algebraically with an exponent around 1.84 after linear extrapolation. For comparison, $P_{xx}(r)$ in the GCE at $M=20000$ decays with an exponent around 0.08. {The extrapolation is not performed for the GCE results because of the convergence limited by the magnitudes of available $M$s. 
%$Y$-bond pair-pair correlations in the GCE are significantly alternating between even and odd legs due to the PDW component and are therefore not shown. In contrast,  %The results in the GCE decay slower than those in the CE for the same reason mentioned earlier for Fig.~\ref{tm02}(c2). 
The exact power-law exponent for d-wave SC correlations should lie between 0.08 and 1.84}. %Nearest-neighbor spin bond orders $-\langle \hat{\boldsymbol{S}}_i\cdot \hat{\boldsymbol{S}}_{\boldsymbol{i+e_\alpha}}\rangle$ in the central region, subtracting the anisotropic $x$- and $y$-bond mean values $\bar{S}^b_x=0.1852$ and $\bar{S}^b_y=0.2215$..
}
\label{PDW}
\end{figure}

 Now we investigate the length scales of Cooper pairs, charge and spin correlations in the pure $t\text-J$ model. 
 As shown in Fig.~\ref{t00_corr}(a), the longer-distance pairing order $\tilde{\Delta}({r})=\langle \hat{c}_{(x_0,y_0),\uparrow}\hat{c}_{(x_0+r,y_0),\downarrow}-\hat{c}_{(x_0,y_0),\downarrow}\hat{c}_{(x_0+r,y_0),\uparrow}\rangle/\sqrt{2}$ decreases exponentially with the distance $r$ between two paired electrons,  characterized by an effective Cooper pair size or coherence length $\xi_P=4.69$.
 Similar small sizes $\xi_P=2.44$ and $2.01$ are also found for $t_2/t_1=\mp 0.2$, respectively~\cite{SM}. 
 %, followed by $t_2/t_1=-0.2$ and then $t_2/t_1=0.2$.
 Meanwhile, the single-particle Green’s function $G(r) = \langle \sum_{\sigma} \hat{c}^{\dagger}_{\sigma }(x_0,y_0) \hat{c}_{\sigma }(x_0+r,y_0) \rangle$,  spin correlation $S(r)=\langle {\hat{\mathbf{S}}}(x_0,y_0) \cdot \hat{\mathbf{S}}(x_0+r,y_0) \rangle$ and 
 charge density correlation $D(r) = \langle {\hat{n} }(x_0,y_0) {\hat{n} }(x_0+r,y_0) \rangle - \langle {\hat{n} }(x_0,y_0) \rangle \langle {\hat{n} }(x_0+r,y_0) \rangle$  can all be described by exponential decays with very short correlation lengths %\deleted{comparable to the Cooper pair size $\xi_P$ }
 (see Fig.~\ref{t00_corr}(b-d)). 
Short-range spin correlations are generally observed for all the SC phases (see SM~\cite{SM}  Fig.~\ref{tpm02_SM}), similar to the previous identified SC phase by doping the spin liquid or spin-liquid like states~\cite{jiang_high_2021,jiang_superconducting_2023},  {in contrast with the long-range spin density waves intertwined with CDW around 1/8 doping reported in the pure Hubbard model~\cite{simons_absence_2020,xu_coexistence_2024} and the extended $t-J$ model on both the electron- and hole-doped sides using a relatively small bond dimension  and spin U(1) symmetry~\cite{jiang_ground-state_2021}. The short-range spin configurations around 1/8 doping in the SC-II and Uni-SC phases are found to be antiferromagnetic (see Fig. S6~\cite{SM}). However, the antiferromagnetic spin configurations exhibit $\pi$-phase shift between neighboring charge stripes in the SC-I phase of the pure t-J model (see Fig.~\ref{t00_corr}(c)), consistent with intertwined short-range  spin and charge stripes.} The coexistence of uniform antiferromagnetism and d-wave SC has been reported at a range of electron doping in large $L_y=8$ systems previously~\cite{jiang_ground-state_2021}. However, we find (see SM Sec. E~\cite{SM}) that the antiferromagnetic spin moment drops significantly to near zero and is expected to vanish with the increase of U(1) bond dimensions, consistent with the short-range spin correlations found in SU(2) simulations (see Fig. S6(b3)~\cite{SM}), whereas the SC order remains stable.%  This indicates the consistency between DMRG simulations imposing different spin symmetries as long as sufficiently large bond dimensions are reached.
%simple kinetic-energy-driven mechanism for the emergence of 
%SC through developing spin-liquid like resonating singlets.
%of spin liquid background: the NN hopping frustrates the AFM order and turns the spin background into a short-range spin liquid.

\begin{figure}[!htbp]
\centering
\includegraphics[width=1.0\linewidth]{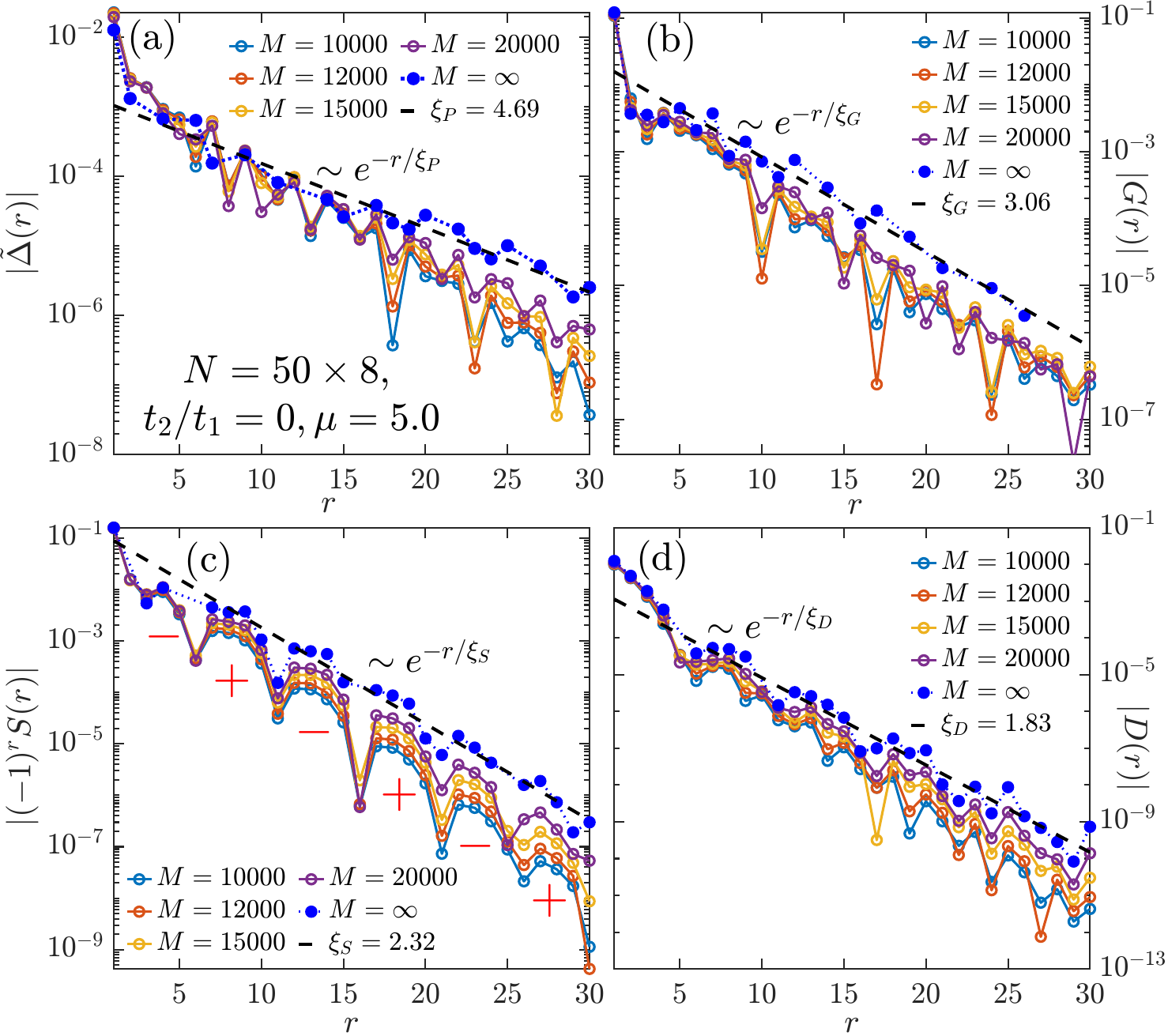}
\caption{Cooper pair  and spin/charge correlation lengths in the SC-I phase at $t_2/t_1=0$ and $\mu=5.0$.   (a) Non-local pairing $\tilde{\Delta}(r)$ versus pair-distance $r$ along $x$-direction 
for different bond dimension $M$s and its extrapolation to infinite $M$, giving  an exponential decay with a coherence length $\xi_P=4.69$
representing the average Cooper-pair size. (b-d) are similar plots for single-particle Green's function, spin and charge-density correlations, with fitted correlation lengths.  {The sign of antiferromagnetic spin correlation $(-1)^rS(r)$ is labeled for each stripe in (c) and it is alternating between neighboring stripes.} % $\xi_G=3.06$, $\xi_S=2.32$ and $\xi_D=1.8$.
}
\label{t00_corr}
\end{figure}

\section{Longer-Distance Cooper Pairs} 
We now explore the  nature of the SC phases through imaging the real space distribution of the Cooper pairs.   
We define  the longer-distance pairing orders $\tilde{\Delta}(\boldsymbol{r})\equiv\tilde{\Delta}(x,y)=\langle \hat{c}_{(x_0,y_0),\uparrow}\hat{c}_{(x_0+x,y_0+y),\downarrow}-\hat{c}_{(x_0,y_0),\downarrow}\hat{c}_{(x_0+x,y_0+y),\uparrow}\rangle/\sqrt{2}$ between two points at positions $(x_0,y_0)$ and 
$(x_0+x,y_0+y)$,  which are shown  
in Fig.~\ref{Nonlocal_pair} as circles  whose  areas are proportional to their magnitudes and whose colors denote their signs.  While the NN pairing orders follow  a  d-wave symmetry for all SC phases, 
%further neighboring orders   
% show sign oscillations in space and decay  with the pairing separation. 
%Besides the dominant d-wave component, 
an additional small $s$-wave component can be identified by the same sign and magnitude of the four NNN pairings at $t_2/t_1=\pm 0.2$ as shown in Fig.~\ref{Nonlocal_pair}(a,c). 
 The longer-distance pairing order shows sign oscillations in real space 
 and decays  with the pairing separation.  The length scale for its sign to repeat along $\hat{x}-$direction is identified to be $l_s=5, 6, 3$, for $t_2/t_1=-0.2, 0, 0.2$, respectively. Furthermore, the Cooper pairs at $t_2/t_1=-0.2$ and 0 are more extended than that at $t_2/t_1=0.2$. The relative magnitudes of $l_s$ for different $t_2/t_1$ are consistent with those of $\xi_P$.

A common feature of all these SC states is the development of pairings between the same sublattices at intermediate distances comparable to $l_s$, indicating that such Cooper pairs gain kinetic energy through NN hoppings. Starting from the undoped pure $t\text{-}J$ model, the pre-existing long-range RVB pairs at half-filling are formed between opposite sublattices~\cite{Liang_1988}, following the Marshall sign rule~\cite{marshall1955} and supporting a long-range AFM order. Upon hole doping, as illustrated in Fig.~\ref{Fig_phase_diagram}(b), the NN electron hopping will bring spin singlets  between opposite sublattices into the same sublattice, breaking the Marshall sign rule and frustrating the AFM order.  {
Therefore, the NN electron hopping is a key driving force in turning the pre-existing RVB pairs due to the  NN spin exchange interactions into coherent Cooper pairs.} 
%For the pure $t\text{-}J$ model, we observe a small diagonal pairing component. }
%\deleted{For the pure $t\text{-}J$ model, the longer-distance Cooper pairs are {potentially} developed through breaking the d-wave symmetry of the NN pair order by inducing an additional unidirectional PDW order intertwining with spin  and charge bond orders.}
%As shown in the supplementary materials\cite{SM} (Fig. S5), 
%a strong-weak unit-2 pattern is observed in both the spin and charge bond orders along y, similar to that of the PDW pattern, indicating additional intertwining between SC, spin and charge orders.
%}
After pinning down the SC for the pure $t\text{-}J$ model,  it becomes clear what is the role of NNN hopping: it helps longer-distance Cooper pairs to  become more symmetric % and suppress
 {by suppressing} the nematicity or PDW tendency of the pure $t\text-J$ model.
Furthermore, the distinct signs of the NNN hoppings for hole- and electron-doped $t\text-J$ models result in different length scales $l_s$ for sign oscillations of $\tilde{\Delta}(\boldsymbol{r})$ and larger $l_s$ on the hole-doped side may make the SC harder to emerge
until the system reaches $L_y=8$~\cite{lu_emergent_2024,lu_sign_2023}. For the electron-doped $t\text{-}J$ model, the SC order also   grows stronger  as the system widens from $L_y=6$ to $8$~\cite{SM}.  {Our results  {suggest} a unified 
picture for understanding the global phase diagram of the extended $t\text{-}J$ model. The cooperative  interplay between the NN hopping and the spin exchange interaction  %.   Their cooperative interplay~\cite{weng_phase_1997} is
is the  key driving force %\deleted{mechanism}
for the general emergence of SC around optimal doping in these systems.}%\deleted{consistent with the fact that the pure t-J model itself has a strong SC ground state as illustrated in the phase diagram}

\begin{figure}[!htbp]
\centering
\includegraphics[width=0.9\linewidth]{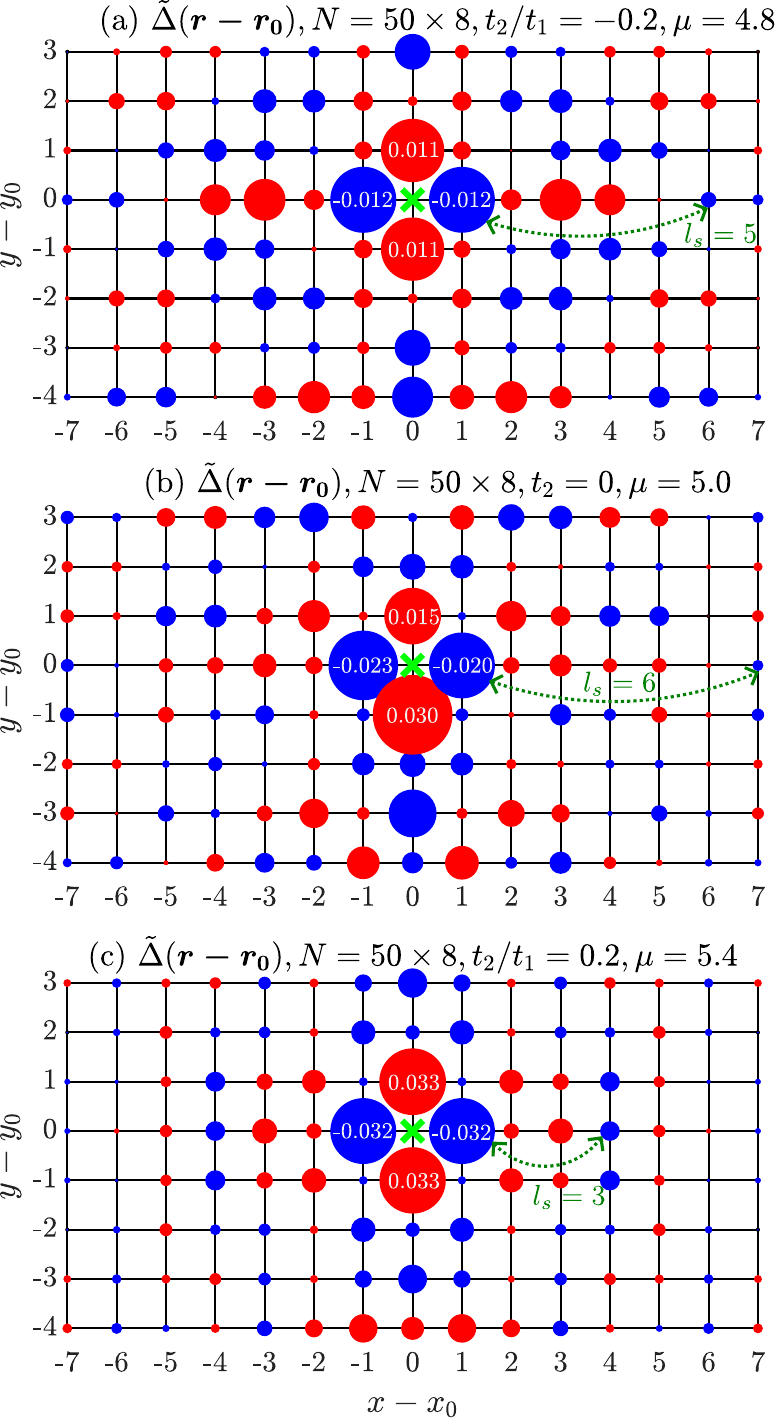}
\caption{Long-distance pairings $\tilde{\Delta}(\boldsymbol{r}-\boldsymbol{r_0})$ imaging real space Cooper pairs.
The reference point $(x_0,y_0)=(25,5)$ for (a-c)  is marked by a green cross. The  areas of the circles are proportional to the magnitudes of $\tilde{\Delta}(\boldsymbol{r}-\boldsymbol{r_0})$ and the red/green color denotes its positive/negative sign. The associated values of nearest-neighbor pairings are given in the figures. Data are obtained under $M=15000$.} 
\label{Nonlocal_pair}
\end{figure}

%\section{Exponentially Decaying Spin and Charge Correlations}

\section{Summary and Outlook}
Through  state-of-the-art DMRG simulations combining grand canonical ensembles~\cite{jiang_ground-state_2021} with a rigorous finite bond dimension analysis, 
we establish a global quantum phase diagram and unveil the  universal emergence of SC 
in the general $t\text{-}J$ model around the optimal hole doping ($0.1\leq \delta \leq 0.2$).
Of particular interest is the pure $t\text{-}J$ model, whose ground state is characterized by  {d-wave SC possibly coexisting with weak
PDW, spin and charge bond orders.}
Our results support the pure $t\text{-}J$ model as a minimal model for understanding unconventional SC in doped Mott insulators, a conclusion of great importance as a d-wave SC ground state has never been observed before in unbiased numerical simulations of the pure $t\text{-}J$ model~\cite{jiang_ground-state_2021,lu_emergent_2024}. 
Beyond the pure $t\text-J$ model, we establish  the  uniform  SC phase in both hole- and electron-doped regimes, where SC grows stronger~\cite{SM} or only emerges~\cite{lu_emergent_2024} 
in wider systems ($L_y=8$), suggesting SC in the 2D limit. 
%\deleted{The necessity of large bond dimensions  and system sizes for the hole-doped SC phase to emerge is due to the stronger competition between kinetic and exchange energies in the presence of additional Berry phase from the negative NNN hopping~\cite{lu_sign_2023}, which leads to spatially larger Cooper pairs with larger length scale of sign oscillations  (Fig.~\ref{Nonlocal_pair})}.}
 {A key driving force for the SC in the $t\text-J$ model
is identified here as the frustrating NN hopping that transforms the  spin singlets  into  phase-coherent Cooper pairs upon increasing doping.  It is important to further identify the role of spin fluctuations, interplay between hoppings and spin correlations,  and the precise mechanism of SC in the future theoretical studies. }

  %Given the observation of prevalent SC in the extended  $t\text{-}J$ model in this work, it is natural to ask if one can also establish a unified mechanism for the unconventional  SC in the one-band Hubbard model~\cite{Jiang_Science_2019,ponsioen_period_2019,simons_absence_2020, qin_hubbard_2022,jiang_ground-state_2024, xu2023coexistence,zhang_frustration-induced_2023} for cuprates.

Ground-state SC has also been identified in the Hubbard model with NNN hopping using both DMRG and 
auxiliary field quantum Monte Carlo methods in the grand canonical finite-size systems in
the presence of a global pairing  field~\cite{xu2023coexistence}.
While their results demonstrate the emergence of SC in both electron- and hole-doped regimes, consistent with our results, the  detailed nature  of Cooper pairs is not known  there, and thus we believe our work will stimulate more extensive exploration of the Cooper pair structure in the extended Hubbard model. % to identify a unified mechanism of unconventional SC. 
Furthermore, it would be interesting to reexamine the pure Hubbard model ($t_2=0$) to understand if the absence of SC in  numerical studies presented
so far~\cite{simons_absence_2020, xu2023coexistence, jiang_ground-state_2024} is related to finite-size and/or low entanglement effects (favoring conventional charge or spin stripes), which will be addressed in the future.

  \section{Method}
 {We adopt the density matrix renormalization group (DMRG)~\cite{white_density_1992} method using its two-site sweep algorithm to study the ground state of the extended $t-J$ model on an eight-leg cylinder with a length up to 50 lattice spacings. DMRG has been established as a highly accurate tool for solving the ground state of quasi-one-dimensional strongly correlated systems. It operates on matrix product states, where the size of the matrices is called bond dimension and controls the Hilbert space dimension of the left/right blocks under bipartition. We impose spin $SU(2)$ symmetry in the matrix product state structures~\cite{mcculloch_non-abelian_2002, gong_robust_2021} and keep large bond dimensions up to $M=15000-20000$ (equivalent to about 45000-60000 U(1) states)  for accurate results, which gives a Hilbert space truncation error smaller or around $3.0\times 10^{-5}$. Simulations at such large bond dimensions are performed using 24 CPU cores, 400 GB of memory and 1 TB of storage for nearly one year. Calculations imposing spin SU(2) and U(1) symmetries are shown to give consistent results (see Fig. S2 in the SM~\cite{SM}). We also test different random initial states to make sure that the calculations do not get stuck in a local minimum.}
 
We consider the grand canonical ensemble (GCE) where spontaneous symmetry breaking of the particle number conservation is allowed~\cite{jiang_ground-state_2021},
 and tune the average electron number by the chemical potential $\mu$. The average doping level is defined as $\delta=1-\langle\sum_i\hat{n}_i\rangle/N$. In the GCE simulations~\cite{jiang_ground-state_2021,SM}, %%with the development of SC order, 
the nonzero  pairing orders (Eq.~\ref{Pairing_order}) detect the spontaneous $U(1)$ symmetry breaking: a low-energy ``natural ground state''~\cite{tasaki_long-range_2019, Jiang_2012}
 with particle number fluctuations can be selected by DMRG algorithm to reduce long-range  entanglement  and Hilbert space fragmentation, effectively lowering the system energy at finite bond dimensions.
 The ``natural ground state''  is a superposition of  low-energy  eigenstates (Anderson Tower of states)~\cite{tasaki_long-range_2019} for a system with the tendency to spontaneous symmetry breaking,  and it becomes near degenerate with the true ground state for large systems. {In contrast, the canonical ensemble (CE) simulations target the symmetric exact ground state and the SC order is examined through the pair-pair correlation functions, which tend to converge slowly with the bond dimension~\cite{jiang_ground-state_2021}.} Thus, the GCE simulations provides a {more efficient} way to detect SC. 

 \section{Code and Data Availability}
 The data for all the figures and the code that produced them are available in the GitHub repository (\href{https://github.com/DongNingSheng/Grand_SQL_tJ/tree/main}{https://github.com/DongNingSheng/Grand\_SQL\_tJ/tree/main}).

\section{Acknowledgement}
 We thank Leon Balents, S. S. Gong, Xin Lu,  H. C. Jiang, Shengtao Jiang, S. A. Kivelson, S. R. White,  Z. Y. Weng, and Wei Zhu for stimulating discussions. This work was supported by the U.S.
Department of Energy, Office of Basic Energy Sciences
under Grant No. DE-FG02-06ER46305 (FC, DNS) for computational study, and National Science Foundation (NSF)
Princeton Center for Complex Materials, a Materials Research Science and Engineering Center DMR-2011750 (FDMH).
D.N.S. also acknowledges partial support from NSF Partnership in Research and Education in
Materials DMR-1828019 for  her travel to Princeton.

\clearpage

\appendix
\widetext
\begin{center}
	\textbf{\large Supplemental Materials}
\end{center}

\vspace{1mm}

\renewcommand\thefigure{ S\arabic{figure}}
\renewcommand\theequation{ S\arabic{equation}}
\renewcommand{\thesubsection}{\Alph{subsection}}

\setcounter{figure}{0} 
\setcounter{equation}{0}  

In the Supplemental Materials, we provide additional numerical results to support the conclusions made in the main text. 
In Sec.~\ref{SC_evol} we show evolution  of pairing orders and pair-pair correlation functions with the bond dimension for six- and eight-leg cylinders by canonical and grand canonical 
simulations with spin $SU(2)$ or $U(1)$ symmetries, establishing the reliability of our simulations. In Sec.~\ref{Supp_convergence}, we provide an example of the finite bond dimension scaling of the ground state energy to demonstrate the numerical convergence of DMRG results. Furthermore,  we demonstrate the equivalence of  spin $SU(2)$ and $U(1)$ simulations in capturing long-range SC pairing orders. {In Sec.~\ref{CDW_SC}, we demonstrate the growth of SC and the suppression of CDW as the bond dimension increases for the hole-doped t-J model. In Sec.~\ref{Supp_pure_t-J_model}, we show pairing correlations of the SC-I phase and demonstrate the weak charge-bond order.} In Sec.~\ref{Supp_corr}, we show the exponentially decaying long-range pairing orders, single-particle Green's function, spin and density correlations for $t_2/t_1=\pm 0.2$. 
{In Sec.~\ref{Spin_U1}, we show the evolution of the spin and SC orders with the increase of the bond dimension in large-scale grand canonical spin-$U(1)$ DMRG simulations.}

%%%%%%%%%%%%%%%%%%%%%%%%%%%%%%%%%%%%%%%%%%%%%%%%%%%%%%%%%%%
\section{Evolution of SC Order and Pairing Correlations in DMRG Simulations}
\label{SC_evol}
As discussed in the main text, the DMRG simulations allow spontaneous symmetry breaking~\cite{jiang_ground-state_2021}, through targeting 
the ``natural ground state''  as a superposition of  low-energy states (Anderson Tower of states)~\cite{tasaki_long-range_2019},
  which becomes  near degenerate with the exact ground state for large systems.
In this section, we provide systematical comparisons between canonical and grand canonical DMRG results with spin $SU(2)$ or $U(1)$ symmetries.
In Fig.~\ref{SC_M}(a), we show how the SC order evolves with the bond dimension $M$ (or $m$) in the grand canonical finite-size  $SU(2)$ (or $U(1)$) DMRG simulations at a fixed chemical potential $\mu=5.4$ and $t_2/t_1=0.2$. ITensor library~\cite{itensor} has been used for the $U(1$) simulations. On the six-leg systems, the SC order sharply increases between  $m=1000$ and 3000. 
After reaching a maximum value, $\Delta_y$ remains strong with the increase of the bond dimension, and such a behavior becomes
more robust for a larger system with $N=50\times 8$, as demonstrated by the larger maximum value of $\Delta_y$ and slower decay with $M$. 
In the whole range of bond dimensions,  the obtained $\Delta_y$ and  ground state energy for $SU(2) $ and $U(1)$ DMRGs
agree with each other as long as we take $M\sim m/3$, demonstrating the higher efficiency of SU(2) simulations. 
 The development of SC order, its plateau behaviour and increased stability with the growing of system sizes reveal the robustness of the SC state in the thermodynamic limit.
 This is similar to the detection of the N\'eel order in the thermodynamic limit~\cite{jiang_ground-state_2021} by the finite-size spin-$U(1)$ DMRG simulations of the antiferromagnetic square-lattice Heisenberg model.

We further compare in Fig.~\ref{SC_M}(b, c) the pairing correlation obtained by both grand canonical and canonical finite-size simulations, respectively. For smaller $M$, the grand canonical simulation will first identify mean-field-like bulk orders, which correspond to a nearly flat pairing correlations at long distances.   As $M$ increases,  the system can take into account more and more quantum fluctuations around the mean-field orders, such that the pairing correlations scales down with $M$ and established a power-law decay with an exponent around 0.62,  indicating a quasi-long-range SC order for the 6-leg ladder systems. For the canonical DMRG calculation using a fixed particle number (Fig.~\ref{SC_M}(c)), the SC pairing correlations at longer distances
are being suppressed for a small bond dimension $M_c$, and a larger $M_c$ is required for their development. Thus
the pairing correlations scale up with $M_c$ and reach a very similar power-law decay with an exponent around 0.55 after the extrapolation. 
The similar exponents obtained by both grand canonical and canonical DMRGs demonstrate the equivalence of the two methods and the good convergence reached for $L_y=6$ systems.  For $L_y=8$ systems, we obtain reliable results for bulk order and qualitatively accurate behavior for pairing correlations, but bond dimensions much larger than currently accessible ones are needed to obtain accurate power-law exponents.  Thus, for a large system, grand canonical DMRG is a proper way to detect ground state SC orders, approximating 
physics in the thermodynamic limit as long as such orders are robust with the increase of the bond dimensions and system sizes.

{To demonstrate that the spin-SU(2) symmetry imposed does not bias SC over CDW on the hole-doped side ($t_2/t_1<0$), in Fig.~\ref{U1_SU2} we compare the spin-SU(2) results with the spin-U(1) ones on a six-leg cylinder at $t_2/t_1=-0.2$ and $\delta=1/32$ in the canonical ensemble. This system has been determined to be in the SC+CDW phase according to Ref.~\cite{lu_emergent_2024} based on SU(2) DMRG simulations. Consistency is found for both the charge density profiles and the decay rates of SC correlations between simulations using two different spin symmetries. Note that this comparison is hard to be made in the GCE because such a low hole doping is very sensitive to the chemical potential and bond dimension.
In eight-leg systems, we estimate that a U(1) bond dimension $m\approx 30,000$ is required to see SC at $t_2/t_1=-0.2$, which is beyond our capability.}

\begin{figure*}[!htbp]
\centering
\includegraphics[width=1\linewidth]{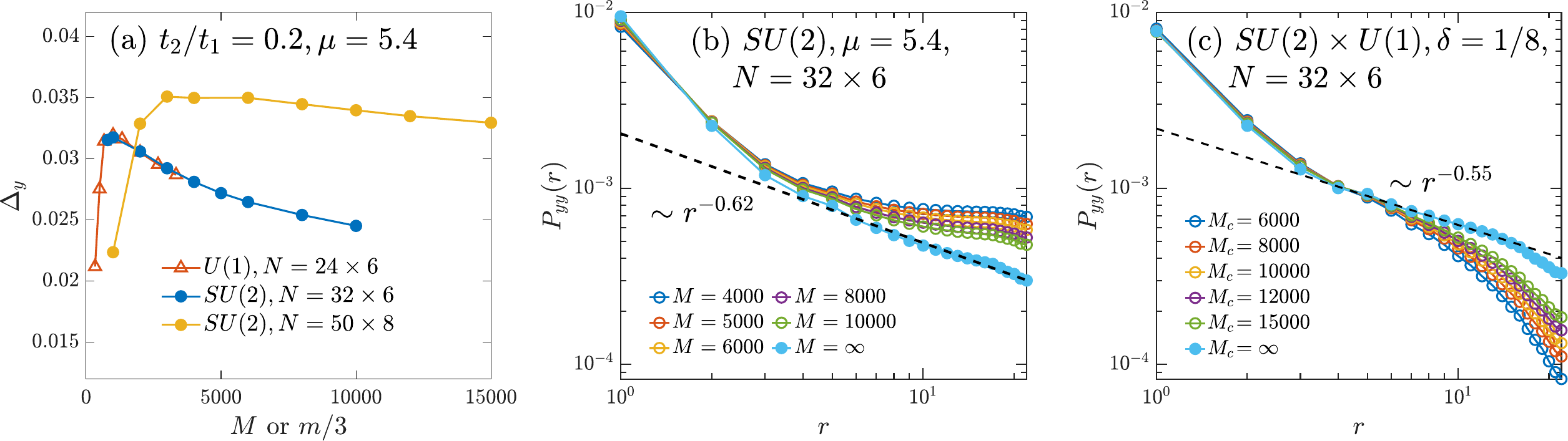}
\caption{(a) $t_2/t_1=0.2$ and $\mu=5.4$. The change of bulk pairing order parameters with the bond dimensions. The $SU(2)$ bond dimension $M$ is counting the number of spin $SU(2)$ multiplets, which  corresponds to about 3 times of the $U(1)$ bond dimension $m$. Both six-leg and eight-leg systems are studied, and the latter show larger SC orders and more robust  plateau for $\Delta_y$ with respect to  $M$, indicating SC in the thermodynamic limit. (b) Pairing correlations $P_{yy}(r)$ at different $M$s obtained by spin $SU(2)$ DMRG for grand canonical systems at fixed $\mu=5.4$ for $t_2/t_1=0.2$ and $N=32\times 6$. A 2nd-order extrapolation to infinite $M$ renders a power-law decay with an exponent around 0.62. (c) Similar plot for a canonical system at doping $\delta=1/8$ and $N=32\times 6$, with the power-law exponent around 0.55.}
\label{SC_M}
\end{figure*}

\begin{figure}
   \includegraphics[width=0.7\textwidth,angle=0]{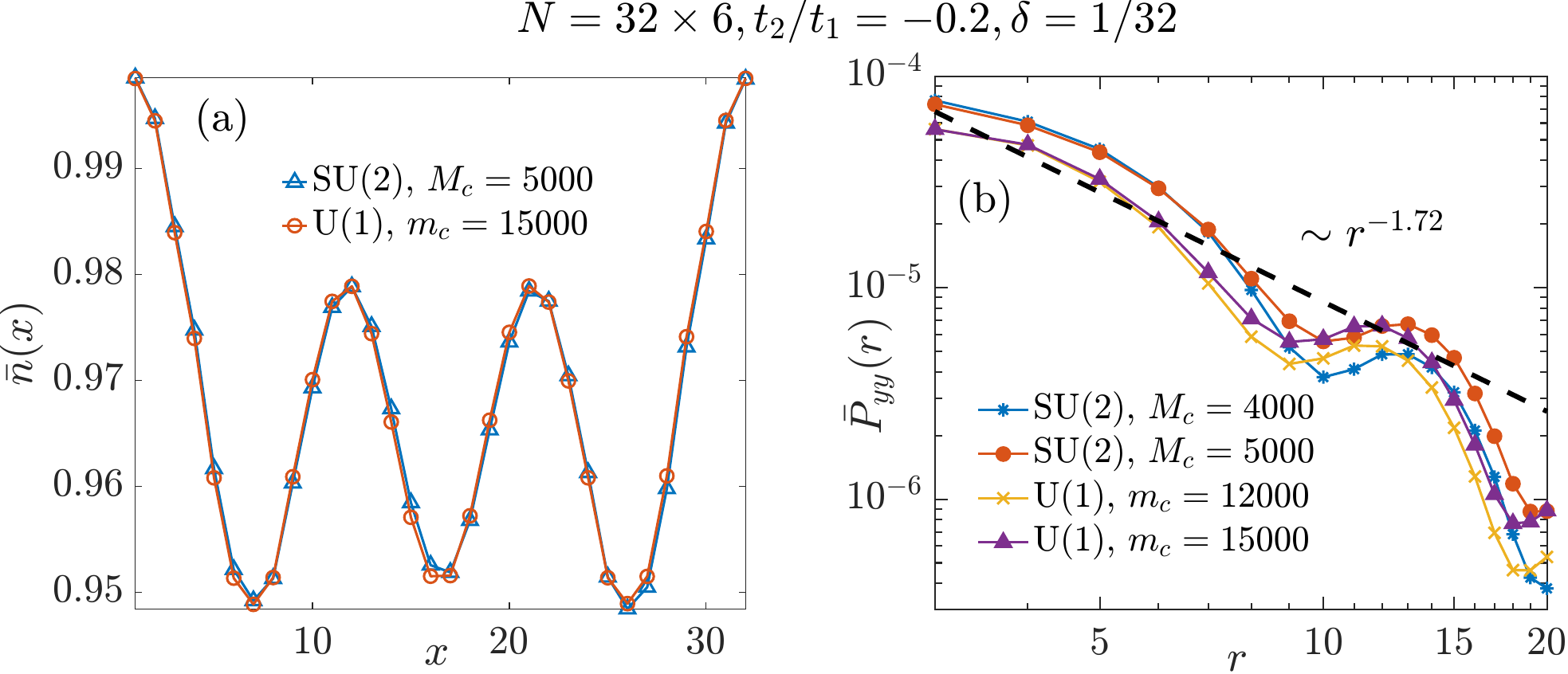}
   \caption{\label{U1_SU2} {
  Spin U(1) and SU(2) DMRG simulations for $N=32\times 6, t_2/t_1=-0.2$ and $\delta=1/32$ in the CE. Rung-averaged density profiles $\bar{n}(x)=\sum_y n(x,y)/L_y$ (a) and pair-pair correlation functions $\bar{P}_{yy}(r)=\sum_{y_0=1}^{L_y} P_{yy}(r;y_0)/L_y$ (b) are consistent between both spin symmetry constraints when $m_c \sim 3M_c$. }}
\end{figure}

\section{DMRG Convergence}
\label{Supp_convergence}
Most of the results presented in the main text are obtained for a system size of $N=50\times8$, and we keep bond dimension upto $M=15000$, which gives a truncation error smaller or around $3.0\times 10^{-5}$. We show in Fig.~\ref{Dr_L6_SM}(a) and (b) the scaling of the ground state energies in terms of the bond dimension for both $L_y=6$ and $8$, which gives the energy density $E/N=-6.590$ and -6.593, respectively. The closeness between energy densities at the largest bond dimension and those extrapolated to infinite $M$ for both systems
demonstrate a good convergence of the  ground state. 

In Fig.~\ref{Dr_L6_SM}(c) and (d), we compare the non-local pairing orders on a six-leg system by both $SU(2)$ and $U(1)$
simulations with $m=M=10000$, and find quantitative agreement between them, indicating a robust structure of the Cooper pair insensitive to the bond dimension or spin symmetry imposed in our simulations.

\begin{figure}[!htbp]
\includegraphics[width=0.8\linewidth]{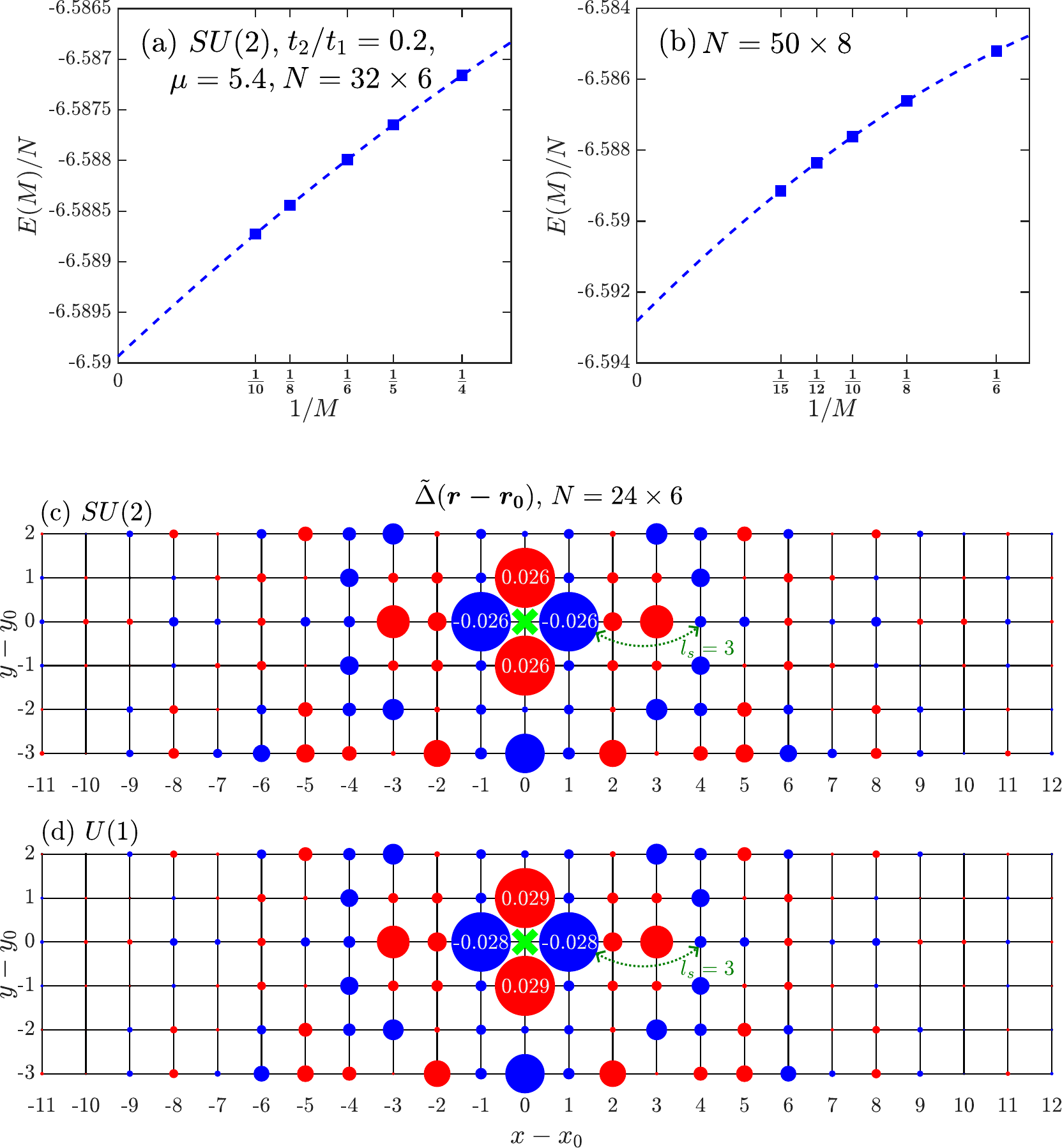}
\caption{$t_2/t_1=0.2, \mu=5.4$. (a) Ground-state energy extrapolation to infinite $M$ for $N=32\times6$ with extrapolated energy  $E(M\to\infty)/N= -6.590$. (b) Similar to (a) for $N=50\times 8$ with  $E(M\to\infty)/N= -6.593$.  (c) and (d) are non-local pairings $\tilde{\Delta}(\boldsymbol{r-r_0})$ imaging Cooper pairs for $N=24\times 6$ obtained by spin $SU(2)$ (with $M=10000$) and spin $U(1)$ (with $m=10000$) DMRG simulations respectively, which are quantitatively similar.}
\label{Dr_L6_SM}
\end{figure}

\section{{Competition between CDW and SC in the Hole-Doped $t-J$ Model}}
\label{CDW_SC}
\begin{figure}
   \includegraphics[width=0.4\textwidth,angle=0]{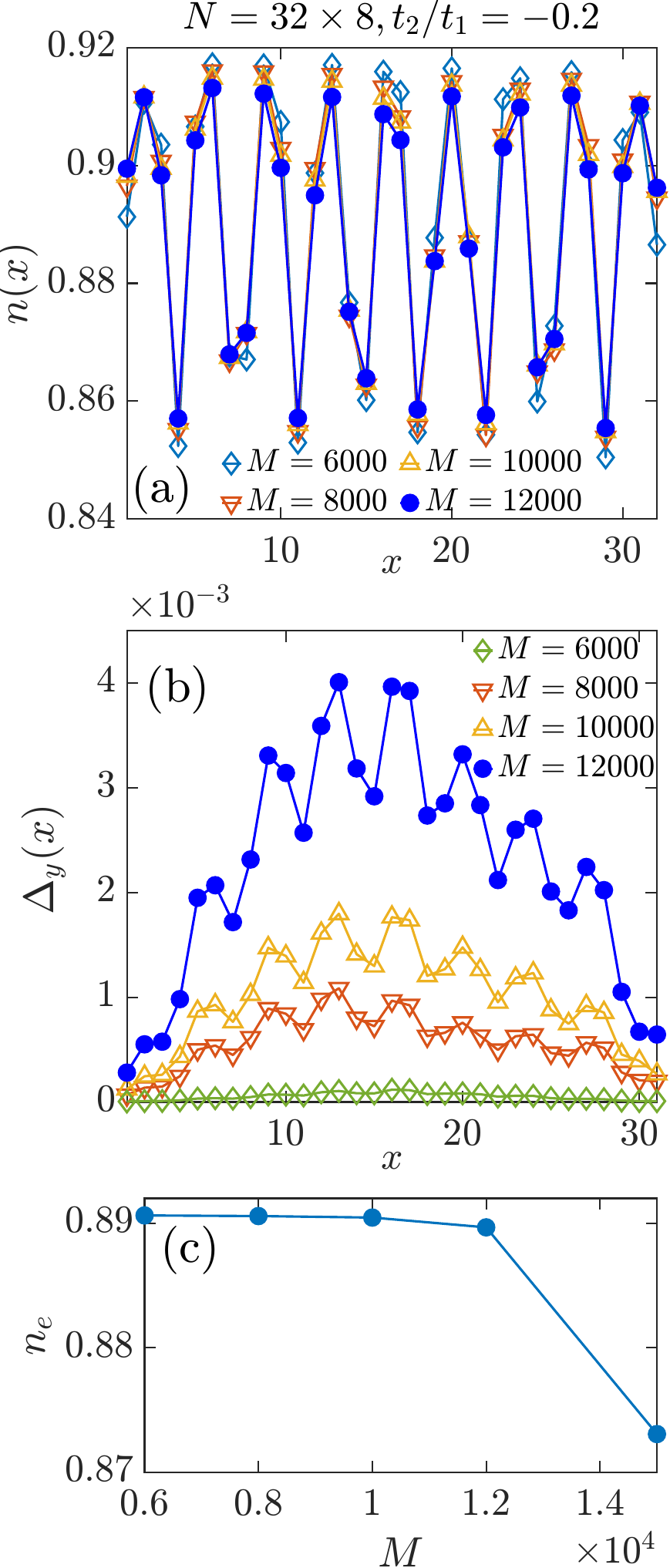}
   \caption{\label{CDW_SC_M}  {
  (a) The weakening of CDW, (b) the strengthening of SC and (c) the invariance of the average electron density $n_e=\sum_i n_i/N$ as the bond dimension increases from 6000 to 12000, while there is a drop of $n_e$ as $M$ increases from 12000 to 15000. $N=32\times 8, t_2/t_1=-0.2$ and $\mu=4.8$.}}
\end{figure}

{In this section, we provide more details of the evolution of CDW, SC and average electron density with the increase of $M$ to support the discussions in Sec. IV in the main text. The rise of SC and the suppression of CDW as $M$ increases from 6000 to 12000 demonstrates the more entangled nature of the SC phase over CDW and the competition between them. The drop of the electron density when $M$ increases from 12000 to 15000 is also a contributing factor to the strengthening of SC shown in Fig.~3(b1) in the main text.}

\section{Spin- and Charge-Bond Orders in the Pure $t\text{-}J$ Model}
\label{Supp_pure_t-J_model}

In the main text, we showed  the pattern of the PDW order.% and its coupling with a spin-bond order.  
 Here we present in Fig.~\ref{Pyy_CBO} the coupled spin and charge bond orders in the SC-I phase.
%The pairing correlation $P_{yy}$ for the pure $t\text{-}J$ model is nearly flat in the middle range in Fig.~\ref{Pyy_CBO}(a) and (b), reflecting the near long-range SC order. The strong-weak alternation in its magnitude along $\hat{y}$ is also exhibited by comparing pairing correlations for $y=1, 2$ in (a) and (b), respectively. In the inset of (b), we show the  mean values of $P_{yy}(x,y=2)$ in the middle 10 sites of the system for different $M$,  which decays with the increase of $M$, indicating a reduced PDW order in the infinite $M$ limit.
A strong-weak unit-2 pattern is observed in both the spin and charge bond orders along $\hat{y}$, similar to that of the PDW in Fig. 4(c) in the main text, indicating additional intertwining between SC, spin and charge orders.

\begin{figure}[!htbp]
\centering
\includegraphics[width=1\linewidth]{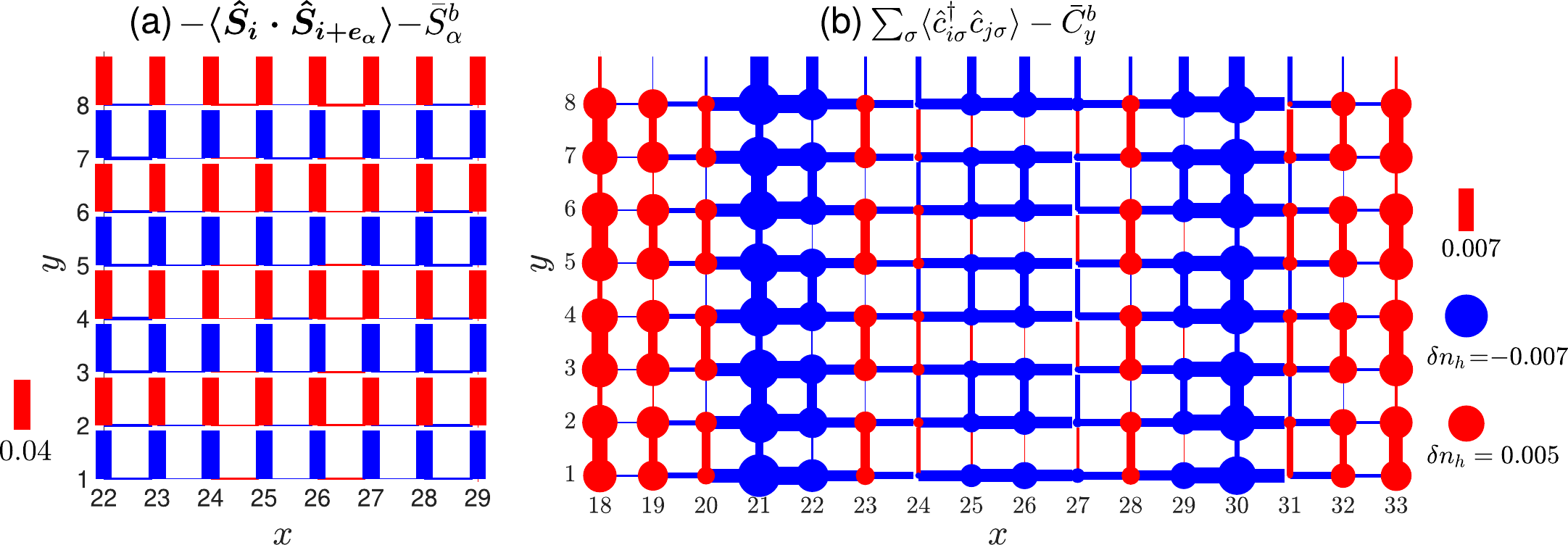}
\caption{$N=50\times 8, t_2/t_1=0$ and $\mu=5.0$. (a) Nearest-neighbor spin bond orders $-\langle \hat{\boldsymbol{S}}_i\cdot \hat{\boldsymbol{S}}_{\boldsymbol{i+e_\alpha}}\rangle$ in the central region, subtracting the anisotropic $x$- and $y$-bond mean values $\bar{S}^b_x=0.185$ and $\bar{S}^b_y=0.221$.
%(a-b) Pairing correlations $P_{yy}(x,y)$  for $y=1,2$  at different bond dimensions. %In the inset of (b), we show the  mean values of $P_{yy}(x,y=2)$ in the middle 10 sites of the system for different $M$, and extrapolate them to infinite $M$ limit. 
(b) Nearest-neighbor charge bond orders $\sum_\sigma\langle\hat{c}_{i\sigma}^\dagger\hat{c}_{j\sigma}\rangle-\bar{C}_y^b$ and hole density variations $\delta n_h(\boldsymbol{r})\equiv n_h(\boldsymbol{r})-\bar{n}_h$, with the subtracted mean values $\bar{C}_y^b\equiv\sum_i\sum_\sigma\langle\hat{c}_{i\sigma}^\dagger\hat{c}_{i+\boldsymbol{e}_y,\sigma}\rangle/N_c=0.095$ and $\bar{n}_h=\sum_i n_h(i)/N_c=0.134$, where $N_c$ is the number of sites in the figure. Results are obtained at $M=12000$}
\label{Pyy_CBO}
\end{figure}

\section{The Spatial Range of Cooper Pairs and the Exponential Decay of Correlations in the Hole- and Electron-Doped $t\text{-}J$ Models}
\label{Supp_corr}

In the main text, we showed various correlations for the pure $t\text-J$ model.  Here we present additional results  in the presence of a small $t_2$.
The long-distance pairing  $\tilde{\Delta}(r) 
=\langle\hat{c}_\uparrow(\boldsymbol{r_0})\hat{c}_\downarrow(\boldsymbol{r_0}+r\boldsymbol{e_x})-\hat{c}_\downarrow(\boldsymbol{r_0})\hat{c}_\uparrow(\boldsymbol{r_0}+r\boldsymbol{e_x})\rangle/\sqrt{2}$, 
single-particle Green’s function $G(r) = \langle \sum_{\sigma} \hat{c}^{\dagger}_{\sigma }(x,y) \hat{c}_{\sigma }(x+r,y) \rangle$, charge density correlation $D(r) = \langle {\hat{n} }(x,y) {\hat{n} }(x+r,y) \rangle - \langle {\hat{n} }(x,y) \rangle \langle {\hat{n} }(x+r,y) \rangle$ and spin correlation $S(r)=\langle {\hat{\mathbf{S}}}(x,y) \cdot \hat{\mathbf{S}}(x+r,y) \rangle$ can all be described by exponential decays with very short correlation lengths (see Fig.~\ref{tpm02_SM}).   {Antiferromagnetic spin patterns are observed in Fig.~\ref{tpm02_SM} (a3), (b3) and (c3)}, and  the short-range spin correlations are consistent with a short-range RVB spin background, driven by the frustrating NN hopping. 

\begin{figure}[!htbp]
\centering
\includegraphics[width=1\linewidth]{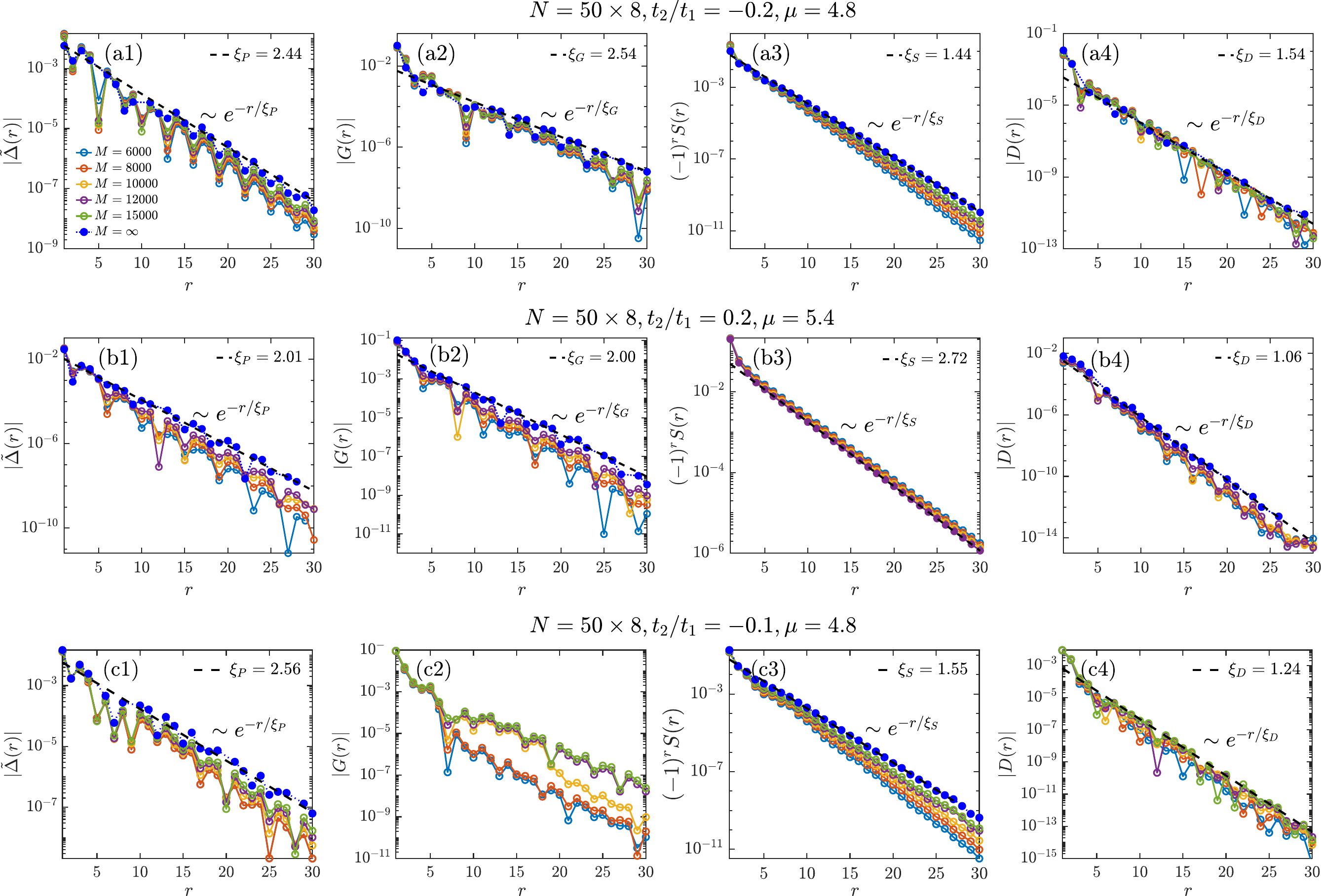}
\caption{(a1-a4) Non-local pairing $\tilde{\Delta}(r)$, single-particle Green's function $G(r)$, antiferromagnetic spin correlation $(-1)^r S(r)$ and density-density correlation $D(r)$ for $t_2/t_1=-0.2$ in the Uni-SC phase obtained at different bond dimension $M$s and their extrapolations to infinite $M$, which are then fitted by exponential decays. (b1-b4) are similar plots for $t_2/t_1=0.2, \mu=5.4$ and $N=50\times 8$ in the Uni-SC phase.  {(c1-c4) are similar plots for $t_2/t_1=-0.1, \mu=4.8$ and $N=50\times 8$ in the SC-II phase. Extrapolations are only done when they are applicable.}
}
\label{tpm02_SM}
\end{figure}

\section{{Vanishing of the AFM Order and the Robustness of the D-wave SC}}
\label{Spin_U1}

{
For a range of doping, the coexistence of  d-wave SC and weak AFM  was observed in Ref.~\cite{jiang_ground-state_2021} based on spin U(1) DMRG simulations on 8-leg systems with the bond dimension $m=4000$. Here in Fig.~\ref{Sz_U1} we show that the spin polarization drops significantly with the increase of $m$ and is expected to vanish around $m=30000$, whereas the SC order remains robust. }

\begin{figure}[!htbp]
\centering
\includegraphics[width=1\linewidth]{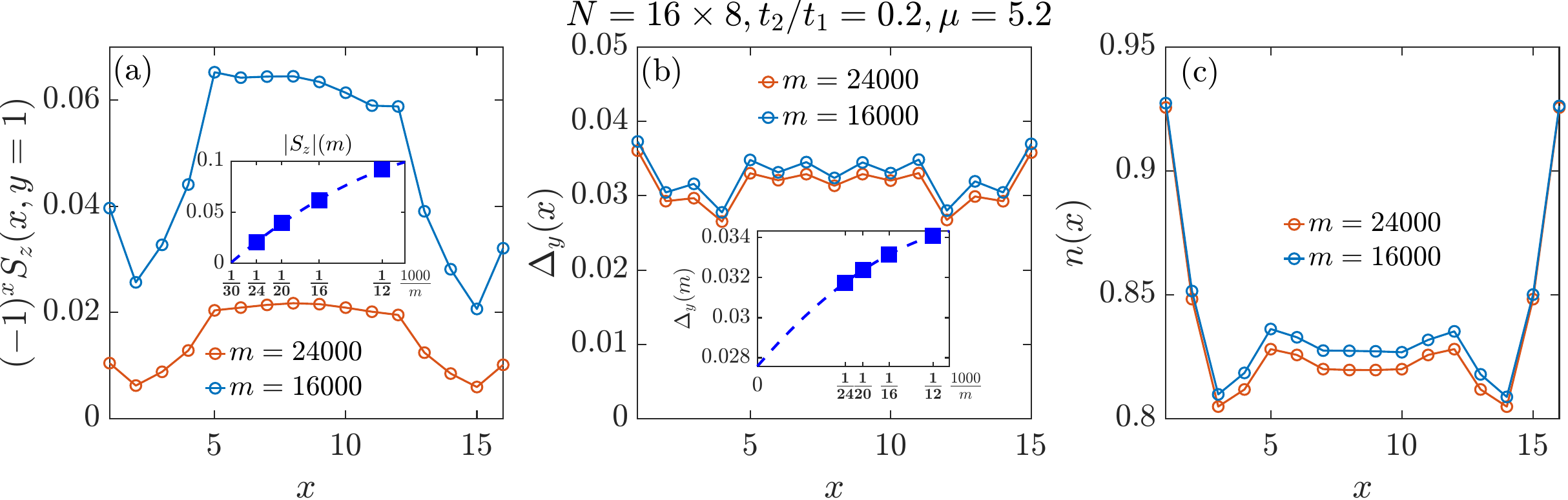}
\caption{{Spin U(1) DMRG simulations for $N=16\times 8, t_2/t_1=0.2$ and $\mu=5.2$ from $m=10000$ to $m=24000$. (a) The antiferromagnetic order is diminishing as $m$ increases and is expected to vanish around $m=30000$; (b) The pairing order $\Delta_y$ is stable with the increase of $m$ and is extrapolated to $\Delta_y(M=\infty)\approx 0.028$; (c) Electron density profiles.}}
\label{Sz_U1}
\end{figure}

%\begin{figure}[!htbp]
%\centering
%\includegraphics[width=1\linewidth]{N32x6, t2=-0.2, mu=5.8.pdf}
%\caption{$N=32\times 6, t_2/t_1=0.2, \mu=5.8, M=12000$. (a) Electron density (b) SC orders at different legs. Ground state energy is $E_g=-1337.9261$. Doping is $\delta=1/27$ }
%\label{tpm02_L6_SM}
%\end{figure}

\end{document}